\documentclass[ reprint ]{revtex4-2}
\usepackage{multirow,eurosym,amssymb,amsfonts,setspace,graphicx,bm,float,amssymb}

\usepackage{color}

\newcommand{\bra}[1]{\langle #1|}
\newcommand{\ket}[1]{|#1\rangle}

\def\be{\begin{equation}}
\def\ee{\end{equation}}

\def\bsplit{\begin{split}}
\def\nsplit{\end{split}}

\begin{document}
\title{Commutation simulator for open quantum dynamics}

\date{\today}

\author{Jaewoo Joo} 
\address{School of Mathematics and Physics, University of Portsmouth, Portsmouth PO1 3QL, UK}

\author{Timothy P. Spiller} 
\address{York Centre for Quantum Technologies, Department of Physics, University of York, York, YO10 5DD, U.K}

\begin{abstract} 
Recent progress in quantum simulation and algorithms has demonstrated a rapid expansion in capabilities. The search continues for new techniques and applications to exploit quantum advantage. Here we propose an innovative method to investigate directly the properties of a time-dependent density operator $\hat{\rho} (t)$. Using generalised quantum commutation simulators, we can directly compute the expectation value of the commutation relation and thus of the rate of change of $\hat{\rho} (t)$. The approach can be utilised as a quantum eigen-vector solver for the von Neumann equation and a decoherence investigator for the Lindblad equation, by using just the statistics of single-qubit measurements. A simple but important example is demonstrated in the single-qubit case and we discuss extension of the method for practical quantum simulation with many qubits, towards investigation of more realistic quantum systems.
\end{abstract}
\maketitle 
{\widetext

A century ago, very early in the development of quantum mechanics, commutation relations emerged in various crucial roles  \cite{Quantum-phys-01,Quantum-phys-02}. Pairs of non-commuting operators (e.g., the position and momentum operators for a particle) describe the complementary nature of their corresponding physical properties, leading to uncertainty relations between these quantities for quantum systems. Commutators also underpin the time evolution of quantum systems, whether this be of general operators in the Heisenberg picture (or the relevant part of the Interaction picture), or the system density operator in the Schr\"odinger picture, where the time dependence resides in the quantum state or density operator. In this latter picture, the quantum state of a system given by $\ket{\psi (t)}$ evolves according to the Schr\"odinger equation (with units where $\hbar = 1$) \cite{Schroedinger01}, given by 
\begin{eqnarray}
&& \hat{\cal H} \ket{\psi (t)} = i {\partial \over \partial t}  \ket{\psi (t)}\;,
\label{Schr01}
\end{eqnarray}
where $\hat{\cal H}$ is the system Hamiltonian. For an initial state defined as $\ket{\psi (0)} = \ket{\psi_0}$ at $t=0$ and a time-independent $\hat{\cal H}$, the evolution from $0$ to $t$ is determined by the unitary operator $\hat{U} (t)$, such that $\ket{\psi (t)} =\hat{U} (t)  \ket{\psi_0} = e^{ -i \hat{ \cal H} t } \ket{\psi_0}$. 

An equivalent alternative description is via the density operator, so defining this as $\hat{\rho} (t) =  \hat{U} (t) \ket{\psi_0} \bra{\psi_0}  (\hat{U}  (t))^{\dag}$ the evolution is given by the von Neumann equation \cite{VNE01} as
\begin{eqnarray}
{d \over dt} \hat{\rho} (t) = i [\hat{\rho}(t) , \hat{\cal H}  ]\,,
\label{von01}
\end{eqnarray}
where the commutation relation between $\hat{x}$ and $\hat{y}$ is given by $[ \hat{x}, \hat{y} ] = \hat{x} \hat{y} - \hat{y} \hat{x}$. 

The form of the von Neumann equation is very interesting because the time-dependence of the system is expressed directly in terms of the commutation relation between the density operator and the Hamiltonian. The density operator approach provides a direct statistical representation  because the diagonal parts of ${d \over dt} \hat{\rho} (t)$ give the rate of change of the system probability density. 
These always correspond to real numbers, which can be measured for the actual physical system, either through repeated measurements on an identically prepared and evolved single pure system, or through measurements on an ensemble of identical systems all equivalently prepared and evolved. 
We refer to these equivalent approaches as an ``ensemble measurement''. The density matrix approach can also be used to incorporate classical uncertainty (lack of knowledge), in addition to quantum superposition, via (finite-entropy) mixtures of pure quantum states. In this work we will use the density operator approach $\hat{\rho} (t)$, both from the perspective of the reversible von Neumann equation (\ref{von01}) but also to provide scope for the inclusion of classical uncertainty and irreversible evolution.

In quantum theory, the irreversibility inherent in open systems---those coupled to additional environment degrees of freedom---can be modelled by modification and addition of noise terms to either the Heisenberg equation for system operators or the Schr\"odinger equation for system states \cite{Peter-RMP}. However, the density matrix approach forms a very important method for investigating the dynamics of open quantum systems, beyond just the Schr\"odinger equation. The Lindblad master equation is a very widely used and applicable example. This commonly describes an open system interacting weakly with its environment, describing the effects of the environment on the system (generally, decoherence mechanisms) using Lindblad operators $ \hat{\cal L}_j$. These operators modify the von Neumann equation (\ref{von01}) to
\begin{eqnarray}
{d \over dt} \hat{\rho} (t) = i \left[ \hat{\rho}(t) , \hat{\cal{H}} \right] + \sum_{j} \left( \hat{\cal L }_j \hat{\rho}(t)\, \hat{\cal L }^{\dag}_j - {1\over 2} \left\{ \hat{\rho} (t)\, , \hat{\cal L }^{\dag}_j \hat{\cal L }_j \right\} \right),
\label{LindEq01}
\end{eqnarray}
where the anti-commutation relation between $\hat{x}$ and $\hat{y}$ is given by $\{ \hat{x}, \hat{y} \} = \hat{x} \hat{y} + \hat{y} \hat{x}$ \cite{Lindblad01,Lindblad02}. In general the Lindblad operators are not Hermitian and act to introduce decoherence to the system, changing its entropy. The particular case of Hermitian Lindblad operators can be used to model quantum measurements, or noisy external source terms in the system Hamiltonian. For $\hat{\cal L}_j = \openone$ ($ \openone$: identity operator), the Lindblad terms disappear and only the unitary term $i \left[ \hat{\rho}(t) , \hat{\cal{H}} \right]$ survives, thus returning to the von Neumann equation and the unitary evolution of a closed quantum system in time.

Quantum algorithms for simulating the Schr\"odinger equation have been developed extensively since the first rigorous idea of quantum simulation in 1996 \cite{SethLloyd96}. However, the preparation of pure states would appear to intrinsically prohibit research progress on the general simulation of mixed density matrices in quantum circuits. The von Neumann equation is normally treated as an equivalent equation to the Schr\"odinger equation, with no decoherence, but containing direct physical interpretation, such as the diagonal elements describing the probability density of the system in the chosen basis. For open quantum systems, several approaches have recently been developed using a vectorized density operator \cite{PRXQuantum22}
or a trace-out of environmental qubits \cite{Cleve17,Childs17,SBen20,deJong21} as the Lindblad-type equations (also named dissipative quantum computation \cite{Eisert11,Cirac09}).

In this work, we propose a novel method to directly compute, or simulate, matrix elements of ${d \over dt} \hat{\rho} (t)$, by measuring expectation values of the commutation relation in the von Neumann equation (\ref{von01}) and the more general Lindblad equation (\ref{LindEq01}). Consider the case where the system of interest comprises $L$ qubits, so the density operator $\hat{\rho} (t)$ can be represented by a $2^L \times 2^L$ matrix. Our approach provides the (diagonal and off-diagonal) matrix elements 
${d \over dt} \rho_{n,m} (t) = \bra{n} {d \over dt} \hat{\rho} (t) \ket{m}$ with $n, m = 0,...,2^L-1$ ranging over a suitable basis of the system. 
So, for example, if we seek the expectation of the rate of state change in time, given by $\bra{\Phi} {d \over dt} \hat{\rho} (t) \ket{\Phi}$ for some chosen reference state $\ket{\Phi}$, we can perform quantum processing to determine this by measuring the expectation value of the commutator in the von Neumann equation (\ref{von01}), given by $i \bra{\Phi} [\hat{\rho}(t) , \hat{\cal H}]\ket{\Phi}$ in the case of closed quantum systems. For the off-diagonal terms, we can compute $i \bra{\Phi} [\hat{\rho}(t) , \hat{\cal H}]\ket{\Phi'}$ by a sum of expectation values given by another controlled-operator gate, with operator $\hat{A}$ for $\ket{\Phi'} = \hat{A} \ket{\Phi}$. For the case of open quantum systems, it is required to perform additional quantum processing to compute the extra Lindblad terms that depend on the $\hat{\cal L }_j$. 

This paper is constructed as follows. In the next section we present the algorithm for simulation of the quantum commutator. Then we present the application of the generalised algorithm to the von Neumann and the Lindblad equations. A specific example is then given, followed by summary and conclusion.
\begin{figure} [b]
\centering
\includegraphics[width=7.5cm,trim=6cm 6cm 6cm 2cm]{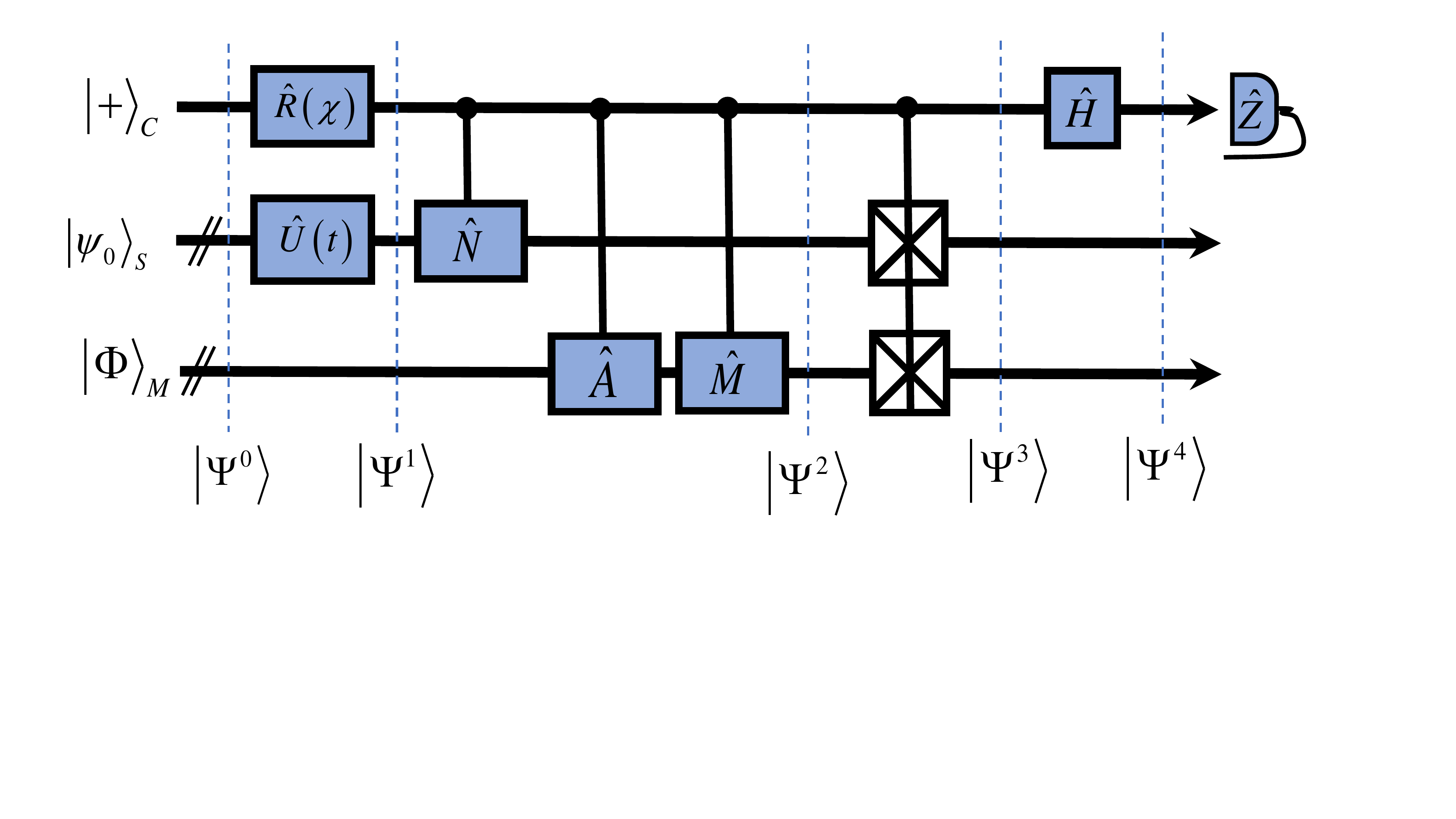}
\caption{Schematic of the generalised quantum circuit to simulate the expectation value of the commutation relation. Three controlled-operators are given by operators $\hat{N}$, $\hat{A}$ and $\hat{M}$ and the total number of qubits required in the simulator is $2L+1$ to describe an $L$-qubit system. The detailed protocol is described in Section \ref{Algorithm:Protocol}.}
\label{fig:01}
\end{figure}

\section*{Algorithm for quantum commutation simulation}
\label{Algorithm}
We first provide the protocol describing the algorithm, followed by a detailed explanation of the quantum commutation simulator. The simulation is built upon the following resources, as employed in Fig.~\ref{fig:01}: The system $S$, assumed to be of dimension $2^L$, or $2^L \times 2^L$ in density matrix form; a separate reference system $M$ of the same size as the system; a separate control qubit $C$. The state of the total system is denoted by $\ket{\Psi}$. With reference to the full quantum circuit shown in  Fig.~\ref{fig:01}, the protocol for the simulation runs as follows.

\subsection*{Protocol}
\label{Algorithm:Protocol}
\begin{enumerate}
\item Initialise the total system state $\ket{\Psi^0}$ as a product of system state $\ket{\psi_0}_S$, reference $\ket{\Phi}_M$ and control qubit $\ket{+}_C$.
\item Perform a single-qubit gate $\hat{R} \,(\chi)$ on $\ket{+}_C$ and system unitary evolution operator $\hat{U} (t)$ on $\ket{\psi_0}_S$ to produce $\ket{\Psi^1}$. 
\item Apply controlled-operator gate $\hat{N}$ between control $C$ and system $S$ as well as two controlled-operator gates $\hat{A}$ and $\hat{M}$ between $C$ and reference $M$ to produce $\ket{\Psi^2}$.
\item Apply a block controlled-SWAP gate from control $C$ between system $S$ and reference $M$ to produce $\ket{\Psi^3}$ \cite{JJ2021}.
\item Apply a Hadamard gate $\hat{H}$ to control $C$ to produce $\ket{\Psi^4}$.
\item Measure a single qubit in $C$ in the Pauli-Z gate (the computational) basis, to obtain the expectation value $\langle \hat{Z} \rangle$. 
\end{enumerate}
}

\subsection*{Quantum commutation simulator}
Using qubit terminology, we first explain the operation of the generalised quantum commutation simulator, shown in Fig.~\ref{fig:01}. Generally, by default, qubits are assumed to be initialised in $\ket{0}$ for a quantum circuit, but here we assume some additional preparation. The control qubit $C$ is prepared in the state $\ket{+}_C = \hat{H} \ket{0}_C$ using a Hadamard gate $\hat{H}$. As shown in Fig.~\ref{fig:01}, there are two $L$-qubit states, for the system $S$ and the reference $M$. For the system $S$, the initial state $\ket{\psi_0}$ is assumed to be created by a suitable prior quantum circuit, specified by the chosen initial conditions of the target problem to be simulated at $t=0$. For the reference $M$, we can simply utilise one of the computational basis states (e.g., $\ket{0}^{\otimes L}$), or any other interesting reference state $\ket{\Phi}$ to be evolved dynamically. Note that only the control qubit $C$ is measured at the end of the process and thus the outcome provides us with the expectation value of a quantum operator for the other degrees of freedom, effectively given by a $2^L \times 2^L$ matrix for each of system $S$ and reference $M$.

For the simulation, a total input quantum state is therefore prepared as $\ket{\Psi^0} = \ket{+}_C \, \ket{\psi_0}_S \, \ket{\Phi}_M$ in Fig.~\ref{fig:01}. In the first step, we apply a single-qubit gate $\hat{R} (\chi)$ on qubit $C$. This is a rotation by $\chi$ around the qubit $Z$-axis with an additional phase, $\hat{R} (\chi) = e^{-i \chi /2} \hat{R}^Z (\chi) $. In parallel, the unitary operator $\hat{U} (t) =  e^{ -i \hat{ \cal H} t }$ is applied to the system $\ket{\psi_0}$, giving the total state as $\ket{\Psi^1} = {1\over \sqrt{2}} ( \ket{0}_C + e^{i \chi} \ket{1}_C)\otimes \ket{\psi (t) }_S \otimes \ket{\Phi}_M$ for $\ket{\psi (t)} = \hat{U} (t) \ket{\psi_0}$.

Next, we apply three controlled-operator gates in order to produce $\ket{\Psi^2}$. These are controlled by qubit $C$ and applied to the system qubits $S$ and to the reference qubits $M$. The operators are represented by $\hat{N}$, $\hat{A}$ and $\hat{M}$ and these controlled operators are each $2^L \times 2^L$ matrices operating on the $L$ reference qubits in either $S$ or $M$. We discuss the roles of these operators and their implementation in more detail in the next section of the paper. 

For the next step of the simulation, to produce $\ket{\Psi^3}$, we apply a block controlled-SWAP gate between the system $S$ and the reference $M$, as shown in Fig.~\ref{fig:01} \cite{JJ2021}. Following this, a Hadamard gate $\hat{H}$ is applied on control qubit $C$, to produce the final (entangled) state of the total system, before measurement 
\begin{eqnarray} 
\ket{\Psi^4} && = {1\over \sqrt{2}} \left( \ket{+}_C \otimes  \ket{\psi (t)}_S \otimes \ket{\Phi}_M 
+ e^{i \chi} \ket{-}_C \otimes \hat{M}\, \hat{A} \,\ket{\Phi}_S \otimes \hat{N} \ket{\psi (t)}_M \right) \; . 
\label{eq:GeneralPsi04}
\end{eqnarray}
Ensemble measurement of the control qubit $C$ in the computational basis will generate the probabilities of the outcomes $\ket{0}_C$ and $\ket{1}_C$, defined respectively as $P_0$ and $P_1$. The expectation value of Pauli operator $\hat{Z}$ is equal to $\left< \Psi^4 \left| \hat{Z} \right| \Psi^4 \right> = P_0 - P_1$.
Since the results are given as a difference of scalar values (details in Appendix \ref{Append:expect01}), we can interchange these to reformulate the expectation value of $\hat{Z}$ for qubit $C$ as
\begin{eqnarray} 
\left< \Psi^4 \left| \hat{Z} \right| \Psi^4 \right> && \equiv \langle \Phi | \hat{Z}^{\chi}_{A} | \Phi \rangle =  {1\over 2} \Big( e^{i \chi} \bra{\Phi}\, \hat{N} \hat{\rho} (t) \hat{M}  \hat{A} \ket{\Phi} + e^{-i \chi}   \bra{\Phi} 
\hat{A}^{\dag} \hat{M}^{\dag}\, \hat{\rho} (t) \, \hat{N}^{\dag} \ket{\Phi} \Big) \; ,
\label{eq:GeneralZ01}
\end{eqnarray}
defining a new quantum operator as
\begin{eqnarray} 
\hat{Z}^{\chi}_{A} = {1\over 2} \Big( e^{i \chi} \hat{N} \hat{\rho} (t) \hat{M}  \hat{A}  + e^{-i \chi}   \hat{A}^{\dag} \hat{M}^{\dag}\, \hat{\rho} (t) \,\hat{N}^{\dag} \Big) \; .  
 \label{eq:GeneralZ01-2}
\end{eqnarray}
Note that this new operator contains actions of the controlled gates $\hat{A}$ and $\hat{M}$, in a manner that depends on the chosen rotation angle $\chi$.

As an example, for an identity $\hat{N} = \hat{A} = \openone$ and $\hat{M}$ being a Hermitian operator $\hat{M}^{\dag} = \hat{M}$, the value of $\chi$ can then determine whether the result delivers the expectation value of the commutation or anti-commutation relation between the time-dependent density matrix $\hat{\rho} (t)$ and the operator $\hat{M}$. These follow from the statistics of single-qubit measurements through
\begin{eqnarray}  
\left\langle {\Phi} \left| \left\{ \hat{\rho} (t), \hat{M} \right\} \right| {\Phi} \right\rangle &=& 2 \, \langle \Phi| \hat{Z}_{\openone}^{0}|\Phi \rangle  \label{eq:GeneralZ0} \\
i  \left\langle {\Phi} \left| \left[ \hat{\rho} (t), \hat{M} \right] \right| {\Phi} \right\rangle &=& 2 \, \langle \Phi| \hat{Z}_{\openone}^{\pi/2}| \Phi \rangle \; . 
\label{eq:GeneralZpi2}
\end{eqnarray}

For $\hat{N} \neq \openone$ and $\hat{A} \neq \openone$, we are further able to utilise the outcome of the expectation value to evaluate both $ \Re \left( \left\langle {\Phi} \left| \hat{N} \hat{\rho} (t) \hat{M}  \right| {\Phi'} \right\rangle \right) = \langle \Phi| \hat{Z}_{A}^{0}|\Phi \rangle$ and $ \Im \left( \left\langle {\Phi} \left| \hat{N} \hat{\rho} (t) \hat{M}  \right| {\Phi'} \right\rangle \right) = - \langle \Phi| \hat{Z}_{A}^{\pi/2}| \Phi \rangle$, for $\ket{\Phi'} = \hat{A} \ket{\Phi}$, where $\Re ()$ and $\Im ()$ represent real and imaginary parts respectively.

\section*{Applications of the quantum commutation simulator}

\subsection*{Dynamics of the von Neumann equation: eigen-state finder}
\label{Dynamics_vN}
We now apply the simulation to the case of the von Neumann equation (\ref{von01}). 
We therefore focus on the case of $\hat{M} = \hat{\cal H}$ as the system Hamiltonian operator. If the Hamiltonian is decomposed in terms of its eigenvalues $\lambda_l$ and eigenvectors $\ket{\lambda_l}$, as
\begin{eqnarray} 
&& \hat{\cal{H}}  = \sum_{l} \lambda_l \ket{\lambda_{l}}\bra{\lambda_{l}}\; ,
\label{GeneralH01}
\end{eqnarray}
then the unitary evolution generated by $\hat{\cal{H}}$ is given by
\begin{eqnarray} 
&& \hat{U}^{tot} (t) = \exp (-i \hat{\cal{H}} t) = \sum_{l} e^{-i \lambda_l t} \ket{\lambda_{l}}\bra{\lambda_{l}} \; .
\label{GeneralUnitary01} 
\end{eqnarray}

The von Neumann equation (\ref{von01}) can then be written in the form
\begin{eqnarray} 
{d \over dt} \hat{\rho} (t) &=& i\, \left[ \hat{\rho} (t), \hat{\cal{H}}\right] = 
i\, \sum_{k \neq l} \, \left(  \lambda_{l}  - \lambda_{k} \right) \, \alpha_k \, \alpha_l^*  \, e^{i (\lambda_l - \lambda_k) t} \, \ket{\lambda_{k}} \bra{\lambda_{l}} \; , 
\label{eigen-vec-find01}
\end{eqnarray}
where $\alpha_k = \bra{\lambda_{k}} {\psi_0} \rangle$ and $\alpha_l^* = \bra{\psi_0} {\lambda_{l}}\rangle$ for
eigenvectors $\ket{\lambda_k}$  and $\ket{\lambda_l}$. In (\ref{eigen-vec-find01}), the matrix of ${d \over dt} \hat{\rho} (t)$ comprises all off-diagonal terms in the  eigenvector representation of the von Neumann equation.

Note that, independent of the reference state $\ket{\Phi}$, the expression $ i\bra{\Phi } [ \hat{\rho} (t), \hat{\cal{H}}]\ket{\Phi} =0$ if either $\alpha_k$ or $\alpha_l$ is zero in each term of the summation. This follows if the initial state $\ket{\psi_0}$ is chosen to equal one of the eigenvectors of $\hat{\cal H}$ in the quantum commutation simulator, rather than being a superposition of two or more eigenvectors with different eigenvalues. Therefore, we can search for eigenvectors of $\hat{\cal H}$ by tuning the parameters in $\ket{\psi_0}$. In general, if $\ket{\psi_0}$ is tuned to include amplitudes of both eigenvectors $\ket{\lambda_k}$  and $\ket{\lambda_l}$, the expectation value $ i\bra{\Phi } [ \hat{\rho} (t), \hat{\cal{H}}]\ket{\Phi}$ oscillates in time with the frequency $ \lambda_{l}  - \lambda_{k}$ and the amplitude $\left(  \lambda_{l}  - \lambda_{k} \right) \, \alpha_k \, \alpha_l^* \langle \Phi \ket{\lambda_k} \langle \lambda_l \ket{\Phi}$. Clearly this is conditional on $\ket{\Phi}$ being of a suitable form so that the amplitude is non-zero.

Interestingly, if $\ket{\Phi}$ is selected as an eigenvector $\ket{\lambda_m}$ of $\hat{ \cal{H}}$,
the expectation value of the right hand side of equation (\ref{eigen-vec-find01}) is always zero, regardless of the time $t$ and the form of $\ket{\psi_0}$, due to the orthogonality of eigenvectors, giving
\begin{eqnarray} 
\left< {\Phi} \left| {d \over dt} \hat{\rho} (t) \right| {\Phi} \right>&=& i\, \left< {\Phi } \left| \,\left[ \hat{\rho} (t), \hat{\cal{H}}\right]\,\right| {\Phi} \right> = 0.
\end{eqnarray}
This result implies that $\ket{\Phi}=\ket{\lambda_m}$ is a stationary state in the unitary dynamics and one of the eigenvectors of the closed system described by $\hat{\cal H}$. 

In addition, the off-diagonal elements of the commutator in the eigenstate basis can be computed. The coherence between the two eigenvectors $\ket{\lambda_k}$  and $\ket{\lambda_l}$ is given by 
$ 
\left< {\lambda_k} \left| \left[ \hat{\rho} (t), \hat{\cal{H}}\right] \right| {\lambda_l} \right> =  \left(  \lambda_{l}  - \lambda_{k} \right) \, \alpha_k \, \alpha_l^*  \, e^{i (\lambda_l - \lambda_k) t} \,
$
for $\langle \lambda_l \ket{\lambda_k} = 0$ with $l \neq k$.  Note that again the amplitude is proportional to the difference of the two eigenvalues $\lambda_{l}  - \lambda_{k}$, as well as this difference setting the frequency of the time-dependent part.

Consider now the specific case where the reference state $\ket{\Phi}$ is chosen as a computational basis state $\ket{n}$ ($n=0,...,2^L-1$ for $L$ qubits), instead of a general computational basis state. Then the expectation value defines the associated probability density rate ${d \over dt} \rho_{n,n} (t) $. Based on the quantum commutation simulator in Fig.~\ref{fig:01}, taking $\hat{N}=\hat{A} = \openone$ and $\hat{M} = \hat{\cal{H}}$ and with $\chi = \pi/2$ in (\ref{eq:GeneralZpi2}), 
we can calculate the diagonal elements of ${d \over dt} \hat{\rho} (t)$ as
\begin{eqnarray} 
{d \over dt} \rho_{n,n} (t) = \left< {n} \left| {d \over dt} \hat{\rho} (t) \right| {n} \right> &&= i \left\langle {n} \left| \left[ \hat{\rho} (t), \hat{\cal{H}} \right] \right| {n} \right\rangle = 2 \langle n | \hat{Z}_{\openone}^{\pi/2} |n \rangle.
\label{eq:vNE03}   
\end{eqnarray}
Thus, if we are able to implement the Hamiltonian operator $\hat{\cal H}$ in the contolled-operator gate of Fig.~\ref{fig:01}, the value of ${d \over dt} \rho_{n,n} (t) $ follows from the statistics of single-qubit measurements of $C$, through $\bra{n} \hat{Z}_{\openone}^{\pi/2} \ket{n}$. 

Moreover, for $\hat{N}= \openone$ and $\hat{M} = \hat{\cal{H}}$ with $\chi = \pi/2$, if $\hat{A}$ is not set as the identity operator but chosen e.g. as the translation operator $\hat{\cal A}$ \cite{JJ2021,JJ2020}, such that $\hat{\cal A}\; \ket{n} = \ket{n+1}$, we can also investigate any off-diagonal matrix elements, such as ${d \over dt} \rho_{n,m} (t) $ for $n \neq m$. Specifically, if we replace $\hat{A}  \rightarrow (\hat{\cal A})^p$, as $p$ actions of the translation operator $\hat{\cal A}$ for the $L$-qubit system ($p=1,...,L-1$), we can compute all the off-diagonal elements of (\ref{eigen-vec-find01}), such as ${d \over dt} \rho_{n,n+p} (t) $.

In any actual implementation of the quantum commutation simulator, one challenge is the construction of the (controlled) operator $\hat{\cal H}$ in the quantum circuit, because the Hamiltonian operator is in general not a unitary operator. 
To resolve this issue, we use the fact that the Hermitian operator $\hat{\cal H}$ can be split into a sum of unitary gates (e.g., Pauli matrices for qubits). Thus, we need to build a decomposition of the Hamiltonian $\hat{\cal{H}} = \sum_{k} c_k \, \hat{\cal U}_k$, with coefficients $c_k$ where each Hamiltonian term $\hat{\cal U}_k$ is represented by a unitary gate. Then, the additivity of commutation relations can be used in equation (\ref{eq:vNE03}) to give $ \left[ \hat{\rho} (t), \hat{\cal{H}} \right] =  \sum_k c_k \left[ \hat{\rho} (t), \,\hat{\cal{U}}_k \right] $. Through this decomposition it is then feasible to simulate the required commutator, via the implementation of controlled-$\,\hat{\cal U}_k$ gates in appropriate quantum circuits.

\subsection*{Dynamics of an open quantum system: decoherence investigator}
The Lindblad equation (\ref{LindEq01}) can describe a decoherence mechanism for an open quantum system weakly coupled with a large environment \cite{Lindblad01,Lindblad02}. As shown in Appendix \ref{Lindblad-derivation}, the derivation of the Lindblad master equation can be described by a series of short time periods $\delta t$. In more detail, the initial density matrix is given in $\hat{\rho}_0 =\ket{\psi_0}\bra{\psi_0}$ and coherently evolves to $\hat{\rho} (\delta t)$ in the period $\delta t$, with the dynamics of the quantum system ${\delta \hat{\rho} (\delta t) \over \delta t }$ given by the von Neumann simulation. During the period between  $\delta t$ and $2\delta t$, the time evolution description of the quantum system is split into two terms: a unitary term with the commutation relation (the same as the von Neumann case) and a sum of the Lindblad terms including the anti-commutation relation \cite{Note1}. With these steps, we are able to rewrite the Lindblad equation at $t=2\delta t$ in the form
\begin{eqnarray}
{\hat{\rho} (2\delta t) - \hat{\rho} (\delta t) \over \delta t} = {\delta \hat{\rho} (2 \delta t) \over \delta t } \approx i \left[ \hat{\rho}(\delta t) , \hat{\cal{H}} \right] + \sum_{j} \left( \hat{\cal L }_j \hat{\rho}(\delta t)\, \hat{\cal L }^{\dag}_j - {1\over 2} \left\{ \hat{\rho} (\delta t)\, , \hat{\cal L }^{\dag}_j \hat{\cal L }_j \right\} \right).
\label{LindEq03}
\end{eqnarray}
The key point is that we can compute all the expectation values of the right hand side terms of equation (\ref{LindEq03}) in the quantum commutation simulator, using both commutation and anti-commutation relations.

In order to obtain the first part of the Lindblad term, let us choose the controlled-operator set $\{ \hat{N}, \hat{M} , \hat{A}\}=\{\hat{\cal L }_j , \hat{\cal L }^{\dag}_j, \openone \} $ in equation (\ref{eq:GeneralZ01}). In the case of $\hat{A} = \openone$, the expectation value of the diagonal elements is given by
\begin{eqnarray}
 \bra{\Phi} \hat{\cal L}_j \, \hat{\rho} (\delta t) \, \hat{\cal L}^{\dag}_j \ket{\Phi} \,&& =  \bra{\Phi} \hat{Z}_{\openone}^{0} \ket{\Phi},
\label{LindEq02-2} 
\end{eqnarray}
which becomes $\bra{n} \hat{\cal L}_j \, \hat{\rho} (\delta t) \, \hat{\cal L}^{\dag}_j \ket{n}$ for $\ket{\Phi}=\ket{n}$ as a computational basis state.
For the off-diagonal elements of the first Lindblad terms, we replace $\hat{A}$ with $(\hat{\cal A})^p$. This enables us to generate the expectation value of any off-diagonal part of the first Lindblad term, which is given by a combination of two different expectation values in the form
\begin{eqnarray}
 \bra{\Phi} \hat{\cal L}_j \, \hat{\rho} (\delta t) \, \hat{\cal L}^{\dag}_j \ket{\Phi'} = \left( \bra{\Phi'} \hat{\cal L}_j \, \hat{\rho} (\delta t) \, \hat{\cal L}^{\dag}_j \ket{\Phi} \right)^{\dag} = \bra{\Phi} \hat{Z}_A^{0} \ket{\Phi} - i\, \bra{\Phi} \hat{Z}_A^{\pi/2} \ket{\Phi}, 
\label{LindEq03-1} 
\end{eqnarray}
where $\langle \Phi \ket{\Phi'} =0$ due to $ \ket{\Phi'} = (\hat{\cal A})^p \ket{\Phi}$. For $\ket{\Phi} = \ket{n}$ 
in the computational basis, $ \bra{n} \hat{\cal L}_j \, \hat{\rho} (\delta t) \, \hat{\cal L}^{\dag}_j \ket{n+p}$ is given by the combination of $\bra{n} \hat{Z}_A^{\chi} \ket{n}$ for $\chi=0,\, \pi/2$.

Lastly, we calculate the anti-commutator component of the second part of the Lindblad term. For the diagonal elements, if $\{ \hat{N}, \hat{M} , \hat{A}\}=\{\hat{\cal L }^{\dag}_j \hat{\cal L }_j, \openone, \openone \}$,
the expectation value of the anti-commutation relation is given by 
\begin{eqnarray}
\left< \Phi \left | \left\{ \hat{\rho} (\delta t), \hat{\cal L }^{\dag}_j \hat{\cal L }_j \right\} \right| \Phi \right>   = 2 \, \langle \Phi| \hat{Z}_{\openone}^{0}|\Phi \rangle.
\label{LindEq04-2}
\end{eqnarray}

For the off-diagonal elements, if we select $\{ \hat{N}, \hat{M} , \hat{A}\}=\{\hat{\cal L }^{\dag}_j \hat{\cal L }_j, \openone, (\hat{\cal A})^p \}$ for $\ket{\Phi '} = \hat{A} \ket{\Phi}$, a combination of two expectation values gives the result
\begin{eqnarray} 
 \left\langle {\Phi} \left| \hat{\cal L }^{\dag}_j \hat{\cal L }_j \, \hat{\rho} (\delta t)  \right| {\Phi'} \right\rangle = \langle \Phi| \hat{Z}_{A}^{0}|\Phi \rangle - i  \langle \Phi| \hat{Z}_{A}^{\pi/2}| \Phi \rangle.
\label{LindEq05}
\end{eqnarray}
Alternatively, if $\{ \hat{N}, \hat{M} , \hat{A}\}=\{ \openone, \hat{\cal L }^{\dag}_j \hat{\cal L }_j , (\hat{\cal A})^p \}$, we obtain
\begin{eqnarray} 
 \left\langle {\Phi} \left| \hat{\rho} (\delta t)  \hat{\cal L }^{\dag}_j \hat{\cal L }_j  \right| {\Phi'} \right\rangle = \langle \Phi| \hat{Z}_{A}^{0}|\Phi \rangle - i  \langle \Phi| \hat{Z}_{A}^{\pi/2}| \Phi \rangle.
\label{LindEq055}
\end{eqnarray}
Thus, adding the results of equations (\ref{LindEq05}) and (\ref{LindEq055}), we obtain the expectation value of the off-diagonal part in the second Lindblad term. Therefore, the expectation values can be computed for all the elements of ${\delta \over \delta t } \hat{\rho}\, (2 \delta t) $ in equation (\ref{LindEq03}).

\section*{1-qubit example}

\subsection*{Theory: amplitude damping for spontaneous emission}
Let us examine the algorithm for the quantum commutation simulator in the case of a single-qubit system under amplitude damping. We consider a general initial system state given by $\ket{\psi_0} = \cos {\theta \over 2} \ket{0} +  e^{i\phi} \sin {\theta \over 2} \ket{1}$, with Hamiltonian $\hat{\cal{H}}  = - {\omega \over 2} \hat{Z}$. This gives the unitary operator $\hat{U} (t) = \exp (-i \hat{\cal{H}} t) =  \exp (i  {\omega \over 2}  \hat{Z} t)$. Writing the evolved density matrix in the form $\hat{\rho} (t) =  \hat{U} (t) \ket{\psi_0}\bra{\psi_0} (\hat{U} (t))^{\dag}$, the right side of the von Neumann equation at time $\delta t$ is given by 
\begin{eqnarray} 
&& i [\hat{\rho}(\delta t) , \hat{\cal{H}} ] =
i {\omega \over 2}  \sin \theta \left( e^{ i \left( \omega \delta t - \phi \right)} \ket{0}\bra{1}  -   e^{i \left( {\phi } -\omega \delta t \right) } \ket{1}\bra{0} \right) \; .
\end{eqnarray}
For the Lindblad equation with a single Lindblad operator $\hat{\cal L }$, we utilse equation (\ref{LindEq03}) at $t=2\delta t$ and obtain a $2 \times 2$ matrix such as 
\begin{eqnarray}
&& {\delta \over \delta t} \hat{\rho} (2 \delta t) = \left(\begin{tabular}{ c c }
$ {\delta \over \delta t}  \rho_{00} (2\delta t) $ & $ {\delta \over \delta t} \rho_{01}(2\delta t) $ \\
$ {\delta \over \delta t}  \rho_{10} (2\delta t) $& $ {\delta \over \delta t} \rho_{11} (2\delta t) $ \\
	\end{tabular}
	\right)   = i [\hat{\rho}(\delta t) , \hat{\cal{H}} ] + \hat{\cal L } \hat{\rho}(\delta t)\, \hat{\cal L }^{\dag} 
- {1\over 2}  \left\{ \hat{\rho} (\delta t), \hat{\cal L }^{\dag} \hat{\cal L } \right\} .
\label{example-Lindblad01}
\end{eqnarray}

\subsection*{Quantum commutation simulation for the von Neumann equation}
For the unitary time evolution of a single-qubit system in the von Neumann equation, we first choose $\hat{N}=\hat{A} = \openone$, $\hat{M}= \hat{Z} = - {2 \over \omega} \hat{\cal H} $ and  $\hat{U} (t) =  \exp (i  {\omega \over 2}  \hat{Z} t)$ in Fig.~\ref{fig:01}. In equation (\ref{eq:vNE03}), the expectation value $\bra{\Phi} \hat{Z}_{\openone}^{\pi/2} \ket{\Phi}$ provides the rate of change of the probability density for $\ket{\Phi}$, so that 
\begin{eqnarray} 
{d \over dt} {\rho}_{00} (t) = - i{\omega \over 2} \left\langle {0} \left| \left[ \hat{\rho} (t), \hat{M} \right] \right| {0} \right\rangle = -\omega \bra{0} \hat{Z}_{\openone}^{\pi/2} \ket{0} = 0 \; , \label{Unitary-rho-dia01} 
\\
{d \over dt} {\rho}_{11} (t) = - i{\omega \over 2} \left\langle {1} \left| \left[ \hat{\rho} (t), \hat{M} \right] \right| {1} \right\rangle = -\omega \bra{1} \hat{Z}_{\openone}^{\pi/2} \ket{1} =0 \; , \label{Unitary-rho-dia02} 
\end{eqnarray}
for the computational basis states $\ket{\Phi} = \ket{0}$ and $\ket{1}$. 
Therefore, regardless of the initial state parameters $\theta$ and $\phi$ in $\ket{\psi_0}$, these expectation values are always zero. This implies (as expected) that the two reference states ($\ket{0}$ and $\ket{1}$) are stationary states in time and the eigenvectors of $\hat{ \cal H}$.

In addition, by choosing $\hat{A} = \hat{X}$, we can compute the off-diagonal elements,  given by the combination of four expectation values (each one for either a real or imaginary part in $ \bra{n}[ \hat{\rho} (t), \hat{M} ] \ket{m} $), such that
\begin{eqnarray} 
&& {d \over dt} {\rho}_{01} (t) = - i{\omega \over 2} \bra{0} [ \hat{\rho} (t), \hat{M} ] \ket{1} = - {\omega \over 2} \left( \bra{0} \hat{Z}_X^{\pi/2} \ket{0} + \bra{1} \hat{Z}_X^{\pi/2} \ket{1} + i \left( \bra{0} \hat{Z}_X^{0} \ket{0} -  \bra{1} \hat{Z}_X^{0} \ket{1} \right) \right) =  i\, { \omega \over 2} e^{ i \left( \omega t - {\phi} \right)} \sin \theta \; ,
\label{Unitary-rho-off01} \\
&& {d \over dt} {\rho}_{10} (t) = - i{\omega \over 2} \bra{1} [ \hat{\rho} (t), \hat{M} ] \ket{0}  =  - {\omega \over 2} \left( \bra{0} \hat{Z}_X^{\pi/2} \ket{0} + \bra{1} \hat{Z}_X^{\pi/2} \ket{1})  - i (\bra{0} \hat{Z}_X^{0} \ket{0} - \bra{1} \hat{Z}_X^{0} \ket{1}) \right) =  -i\, { \omega \over 2} e^{ i \left({\phi}- \omega t \right) } \sin \theta \; .~~~~~~
\label{Unitary-rho-off02} 
\end{eqnarray}

For this case of qubit amplitude damping, Fig.~\ref{fig:02} demonstrates examples of relevant non-zero matrix elements of the Lindblad equation Eq.~(\ref{example-Lindblad01}), for specific parameter choices and as functions of $t$ and $\theta$. In Fig.~\ref{fig:02}(a) we observe that the real part of Eq.~(\ref{Unitary-rho-off01}) oscillates in time for $\theta \neq 0$ and $\pi$. The static cases ($\theta = 0 \;,\; \pi$) identify the eigenvectors $\ket{0}$ and $\ket{1}$ of $\cal{H}$, whereas $\omega$ follows from the period of the oscillation in $\sin (\omega t)$ (with $\omega=-2$ and $\phi=0$ in this example). Thus, the eigenvalue information $\omega$ can be also extracted from the landscape of measurement outcomes.

\subsection*{Quantum commutation simulation for two Lindblad terms}
\label{IV-C}
\begin{figure} [b] 
\centering
\includegraphics[width=16cm,trim=-1cm 0cm -1cm 0cm]{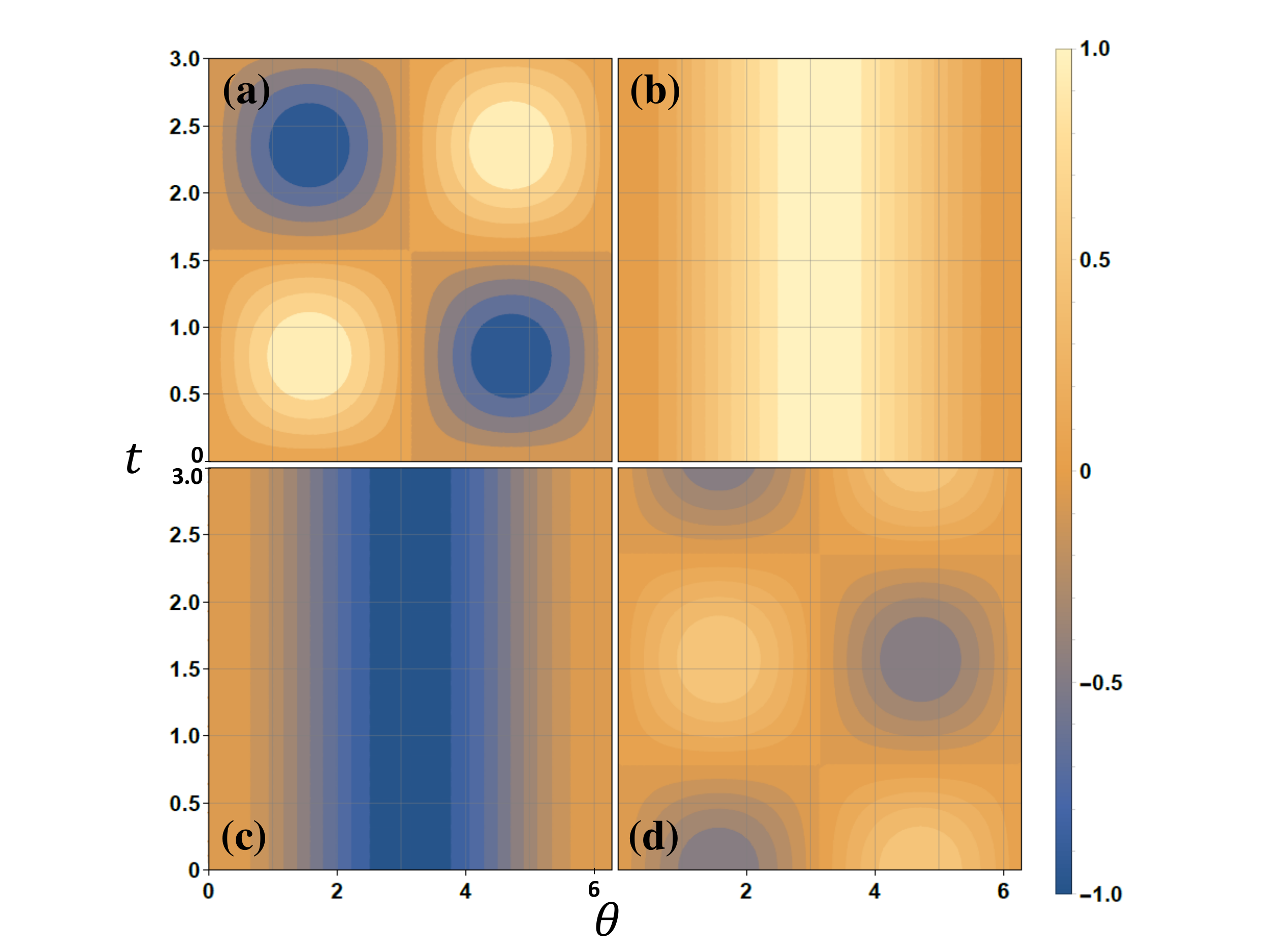}
\caption{ The matrix elements of the Lindblad equation are shown in (a) $ \Re \left[  i \bra{0} [ \hat{\rho} (t), \hat{H} ] \ket{1} \right] =  \bra{0} \hat{Z}_X^{\pi/2} \ket{0} + \bra{1} \hat{Z}_X^{\pi/2} \ket{1}$ from equation (\ref{Unitary-rho-off01}), (b) $\bra{0} \hat{\cal L} \, \hat{\rho} ( \delta t) \, \hat{\cal L}^{\dag} \ket{0}$=$\bra{0}  \hat{Z}_{\openone}^{0}  \ket{0}$  
from equation (\ref{Lind1-dia01}), (c) $-{1 \over 2} \left< 1 \left | \{ \hat{\rho} (\delta t) , \hat{\cal L }^{\dag} \hat{\cal L } \} \right| 1 \right>$=$- \langle 1| \hat{Z}_{\openone}^{0}|1 \rangle$ from equation (\ref{Lind2-dia02}) and (d) $ - \Re \left( \left\langle {0} \left| \hat{\rho} (\delta t)  \hat{\cal L }^{\dag} \hat{\cal L }  \right| {1} \right\rangle  \right)$=$- \langle 0| \hat{Z}_{X}^{0}|0 \rangle$ from equation (\ref{LindEq06}) The parameters are given by $\phi = 0$, $\omega = -2$, $\kappa = 1$, $0 \le \theta \le 2 \pi$ and $0 \le t \le 3$.}
 \label{fig:02}
\end{figure}

For the first Lindblad term in equation (\ref{example-Lindblad01}), we choose that $\hat{N}=\hat{\cal L }$ and $\hat{M}=\hat{\cal L }^{\dag}$ in equation (\ref{eq:GeneralZ01}), to obtain 
\begin{eqnarray}
&&\bra{\Phi} \hat{Z}_A^{0} \ket{\Phi}  =   {1\over 2} \Big( \bra{\Phi}\hat{\cal L}\, \hat{\rho} ( \delta t) \hat{\cal L}^{\dag}  \hat{A} \ket{\Phi} +  \bra{\Phi} \hat{A}^{\dag} \hat{\cal L}\, \hat{\rho} ( \delta t) \hat{\cal L}^{\dag}\, \ket{\Phi} \Big),
\label{LindEq02}\\
&&\bra{\Phi} \hat{Z}_A^{\pi/2} \ket{\Phi}  =   {i\over 2} \Big( \bra{\Phi}\hat{\cal L}\, \hat{\rho} ( \delta t) \hat{\cal L}^{\dag}  \hat{A} \ket{\Phi} -  \bra{\Phi} \hat{A}^{\dag} \hat{\cal L}\, \hat{\rho} ( \delta t) \hat{\cal L}^{\dag}\, \ket{\Phi} \Big).
\end{eqnarray}
For $\hat{A} = \openone$ in equation (\ref{LindEq02-2}), the diagonal part of the first Lindblad term for $\ket{\Phi} = \ket{0}\,, \ket{1}$ is given by
\begin{eqnarray}
&& \bra{0} \hat{\cal L} \, \hat{\rho} ( \delta t) \, \hat{\cal L}^{\dag} \ket{0}= \bra{0}  \hat{Z}_{\openone}^{0}  \ket{0} = \kappa \sin^2 {\theta\over 2}\;, 
\label{Lind1-dia01} \\
&& \bra{1} \hat{\cal L} \, \hat{\rho} ( \delta t) \, \hat{\cal L}^{\dag} \ket{1}= \bra{1}  \hat{Z}_{\openone}^{0}  \ket{1} = 0\;, 
\label{Lind1-dia012} 
\end{eqnarray}
(see Fig.~\ref{fig:02}(b) with $\kappa = 1$) while the off-diagonal parts are given with $\hat{A} = \hat{X}$ and $\ket{\Phi} = \ket{0}$ in equation (\ref{LindEq03-1}), with the result
\begin{eqnarray}
&&  \bra{0} \hat{\cal L} \, \hat{\rho} (t) \, \hat{\cal L}^{\dag} \ket{1} =\left(\bra{1} \hat{\cal L} \, \hat{\rho} (t) \, \hat{\cal L}^{\dag} \ket{0}\right)^{\dag} = \bra{0}  \hat{Z}_X^{0}  \ket{0} - i  \bra{0} \hat{Z}_X^{\pi/2} \ket{0} =0\;.
\label{Lind1-off01}
\end{eqnarray}
Thus, all parts of the expectation values $ \langle \hat{\cal L} \, \hat{\rho} ( \delta t) \, \hat{\cal L}^{\dag} \rangle$ can be computed as above.

Finally, we choose the combination of three controlled operators and the parameter $\chi$ to compute the second Lindblad term in equation (\ref{example-Lindblad01}), based on equations (\ref{LindEq04-2}), (\ref{LindEq05}) and (\ref{LindEq055}). 
The diagonal elements are given with the set $\{ \hat{N}, \hat{M} , \hat{A}\}=\{\hat{\cal L }^{\dag} \hat{\cal L }, \openone, \openone \}$.
As shown in Fig.~\ref{fig:02}(c), the expectation values of the anti-commutator are given by 
\begin{eqnarray}
&& \left< 0 \left | \{ \hat{\rho} (\delta t), \hat{\cal L }^{\dag} \hat{\cal L } \} \right| 0 \right>   = 2 \, \langle 0| \hat{Z}_{\openone}^{0}|0 \rangle = 0 \;,
\label{Lind2-dia01}\\
&& \left< 1 \left | \{ \hat{\rho} (\delta t), \hat{\cal L }^{\dag} \hat{\cal L } \} \right| 1 \right>   = 2 \, \langle 1| \hat{Z}_{\openone}^{0}|1 \rangle = 2 \kappa \sin^2 {\theta \over 2} \; .
\label{Lind2-dia02}
\end{eqnarray}

For $\{ \hat{N}, \hat{M} , \hat{A}\}=\{\hat{\cal L }^{\dag} \hat{\cal L }, \openone, \hat{X} \}$, one of expectation values for off-diagonal elements is given with $\ket{\Phi} = \ket{0}$ by
\begin{eqnarray} 
 \left\langle {0} \left| \hat{\cal L }^{\dag} \hat{\cal L } \, \hat{\rho} (\delta t)  \right| {1} \right\rangle  = \left( \left\langle {1} \left| \hat{\rho} (\delta t)  \hat{\cal L }^{\dag} \hat{\cal L }  \right| {0} \right\rangle \right)^{\dag} = \langle 0| \hat{Z}_{X}^{0}|0 \rangle - i  \langle 0| \hat{Z}_{X}^{\pi/2}| 0 \rangle =0 \, ,
\label{Lind2-off01} 
\end{eqnarray}
while the other becomes for $\{ \hat{N}, \hat{M} , \hat{A}\}=\{ \openone, \hat{\cal L }^{\dag} \hat{\cal L } , \hat{X} \}$ as
\begin{eqnarray} 
 \left\langle {0} \left| \hat{\rho} (\delta t)  \hat{\cal L }^{\dag} \hat{\cal L }  \right| {1} \right\rangle  = \left(  \left\langle {1} \left| \hat{\cal L }^{\dag} \hat{\cal L } \, \hat{\rho} (\delta t)  \right| {0} \right\rangle\right)^{\dag}= \langle 0| \hat{Z}_{X}^{0}|0 \rangle - i  \langle 0| \hat{Z}_{X}^{\pi/2}| 0 \rangle = {\kappa \over 2} e^{i (\omega \delta t \,- \phi)} \sin \theta \, .
\label{LindEq06}
\end{eqnarray}
Thus, we conclude $\left< 0 \left | \left\{ \hat{\rho} (\delta t), \hat{\cal L }^{\dag} \hat{\cal L } \right\} \right| 1 \right> = {\kappa \over 2} e^{i (\omega \delta t \,- \phi)} \sin \theta$ 
and  $\left< 1 \left | \left\{ \hat{\rho} (\delta t), \hat{\cal L }^{\dag} \hat{\cal L } \right\} \right| 0 \right> = {\kappa \over 2} e^{-i (\omega \delta t \,- \phi)} \sin \theta $, and its real part is shown with $\phi = 0$, $\omega = -2$ and $\kappa = 1$ in Fig.~\ref{fig:02}(d).

Therefore, for the single-qubit open quantum system under spontaneous emission, the full Lindblad equation (\ref{example-Lindblad01}) at $t= 2 \delta t$ is given by
\begin{eqnarray}
&& {\delta \over \delta t} \hat{\rho} (2 \delta t)  = \left(\begin{tabular}{ c c }
$ {\delta \over \delta t}  \rho_{00} (2\delta t) $ & $ {\delta \over \delta t} \rho_{01}(2\delta t) $ \\
$ {\delta \over \delta t}  \rho_{10} (2\delta t) $& $ {\delta \over \delta t} \rho_{11} (2\delta t) $ \\
	\end{tabular}
	\right) = \left(
	\begin{tabular}{ c c }
$ \kappa {\rho}_{11} (0)$ & $ \left( i \omega - {\kappa \over 2} \right) {\rho}_{01} (0)$ \\
 $ - \left( i \omega + {\kappa \over 2} \right) {\rho}_{10} (0)$	& $ - \kappa {\rho}_{11} (0)$ \\
	\end{tabular}
	\right) \; .
\label{1-qubit-Lindblad02}
\end{eqnarray}
Let us make various comments at this stage.
\begin{itemize}
    \item The result of equation (\ref{1-qubit-Lindblad02}) is exactly that which we would expect for amplitude damping of a qubit.
    \item We note that the rates of change of the diagonal density matrix elements are no longer zero at $t= 2\delta t$, due to the decoherence from the amplitude damping in the open quantum system.
    \item The most important point to stress is that we have calculated the terms that contribute to equation (\ref{1-qubit-Lindblad02}) in a manner that is amenable to quantum simulations, through various specifically chosen cases of Fig.~\ref{fig:01}. This provides a route for simulation of Lindblad evolution.
    \item We chose an initial ($t=0$) pure state, but an initial mixed state could be simulated as a mixture over a suitable decomposition of pure states.
\end{itemize}

Finally, to fully justify our first comment above, we construct the actual resultant density matrix $\hat{\rho} (t)$. In effect, although we used the terminology of a $t=0$ initial state, what equation (\ref{1-qubit-Lindblad02}) represents is the relationship of the rates of change of the density matrix elements at some given time to the actual matrix elements at that time. Therefore, equation (\ref{1-qubit-Lindblad02}) provides (in this case) the four anticipated first order differential equations for the density matrix, that can be integrated to give the expected result for a qubit with amplitude damping
\begin{eqnarray}
&&  \hat{\rho} ( t)  = \left(\begin{tabular}{ c c }
$   \rho_{00} ( t) $ & $ \rho_{01}(t) $ \\
$  \rho_{10} ( t) $  & $  \rho_{11} (t) $ \\
	\end{tabular}
	\right) = \left(
	\begin{tabular}{ c c } 
$ 1 -{ \rho}_{11} (0)\, \, e^{ - \kappa t} $ & ${\rho}_{01} (0)\, \, e^{\left(i \omega - \kappa/2 \right) t }$ \\
 $ {\rho}_{10} (0) \, e^{\left(-i \omega - \kappa/2 \right) t } $	& ${\rho}_{11} (0)\, e^{ - \kappa t }$ \\
	\end{tabular}
	\right) \; .
\label{1-qubit-Lindblad03}
\end{eqnarray}

Analogous to the controlled-Hamiltonian gate in Section \ref{Dynamics_vN}, the Lindblad operators $\hat{\cal L }$ and $\hat{\cal L }^{\dag}$ are used here. The non-Hermitian operators describing the spontaneous emission with damping parameter $\kappa$ are given by
$\hat{\cal L } = {\sqrt{\kappa} \over 2} \left( \hat{X} +i \hat{Y} \right)$, $\hat{\cal L }^{\dag} = {\sqrt{\kappa} \over 2} \left( \hat{X} -i \hat{Y} \right)$ 
and $\hat{\cal L }^{\dag} \hat{\cal L } = {\kappa \over 2} \left( {\openone} - \hat{Z} \right)$. Then, the implementation of these operators is always feasible in the commutation simulator, given by the decomposition of the operators $ \sum_{ij} c_{jk}\, \hat{\Sigma}_j \, \hat{\rho} (\delta t) \, \hat{\Sigma}_k$ and $\hat{\Sigma}$ denotes the basis for $2 \times 2$ Hermitian matrices, given by the three Pauli operators plus the identity $\openone$. 

\section*{Summary and Remarks}
In summary, we have proposed a new quantum algorithm to simulate the dynamics of open and closed quantum systems in quantum circuits. Two interesting applications of this approach are investigations of: (i) steady states in a closed quantum system, via the von Neumann equation; and (ii) decoherence mechanisms in an open quantum system, via the Lindblad equation. For a large quantum system, the von Neumann method is beneficial for computing a transition rate between two specific quantum states and this result can be also used for the study of its open quantum system. For example, although the sizes of the system qubits state $\ket{\psi_0}$ and the reference qubits state $\ket{\Phi}$ is each given by a $2^L \times 1$ column vector for $L$ qubits, the probability transition rate between two specific states is always given by the expectation value of the off-diagonal elements in a $2 \times 2$ matrix form. Correspondingly, the changes of each state probability follow from the diagonal elements. 

Clearly in general Lindblad evolution is irreversible, with a change in mixture (entropy) of the density matrix. In the simulation approach this changing mixture is introduced because the outputs of different simulations have to be combined, and each of these simulations involve measurements, to compute the expectation values that comprise the various matrix elements of the (rate of change of the) density matrix.

Extensions of the Lindblad equation simulation method have the potential to generate innovative approaches for general purpose master equations. This is because the simulation approach preserves a probabilistic interpretation for the system even for open systems, generating the evolution of the system probabilities from the diagonal elements of a density. The investigation of possible quantum advantage in such open system applications will be an interesting topic for future study, for example to investigate the quantum speed limit of simulating an open quantum system \cite{Saito19,Hasegawa21}.

\section*{Acknowledgements}
This work is supported by Basic Science Research Program through the National Research Foundation of Korea (NRF) funded by the Ministry of Education, Science and Technology (NRF-2021M3H3A1038085).

\section*{Methods}
\subsection*{Expectation value of the control qubit $C$}
\label{Append:expect01}
If we perform single-qubit measurements on $C$ in (\ref{eq:GeneralPsi04}), the probabilities of outcomes $\ket{0}_C$ and $\ket{1}_C$ are given by
\begin{eqnarray} 
&& P_0 = \bra{\Psi^4} 0 \rangle_C \langle 0 \ket{\Psi^4}  
= {1\over 4} \big( 1+  {}_S \bra{\Phi} \hat{A}^{\dag} \, \hat{M}^{\dag} \hat{M} \hat{A} \, \ket{\Phi}_S \otimes {}_M \bra{\psi (t)} \hat{N}^{\dag}\hat{N}\,  \ket{\psi (t)}_M \nonumber \\
&& ~~~~~~~~+ e^{i \chi} {}_S \bra{\psi (t)} \hat{M}\, \hat{A} \, \ket{\Phi}_S \otimes {}_M \bra{\Phi}\hat{N}\,  \ket{\psi (t)}_M + e^{-i \chi}  {}_S \bra{\Phi} \hat{A}^{\dag} \, \hat{M}^{\dag}\,  \ket{\psi (t)}_S \otimes  {}_M \bra{\psi (t)} \hat{N}^{\dag}\, \ket{\Phi}_M \Big),~
 \label{eq:GeneralExpect0} \\
&& P_1 = \bra{\Psi^4} 1 \rangle_C \langle 1 \ket{\Psi^4}  
 = {1\over 4} \big( 1+  {}_S \bra{\Phi} \hat{A}^{\dag}\, \hat{M}^{\dag} \hat{M} \hat{A} \, \ket{\Phi}_S \otimes {}_M \bra{\psi (t)} \hat{N}^{\dag}\hat{N}\,  \ket{\psi (t)}_M \nonumber \\
&& ~~~~~~~~- e^{i \chi} {}_S \bra{\psi (t)} \hat{M}\, \hat{A} \, \ket{\Phi}_S \otimes {}_M \bra{\Phi}\hat{N}\,  \ket{\psi (t)}_M - e^{-i \chi}  {}_S \bra{\Phi} \hat{A}^{\dag} \, \hat{M}^{\dag}\,  \ket{\psi (t)}_S \otimes  {}_M \bra{\psi (t)} \hat{N}^{\dag}\, \ket{\Phi}_M \Big).~
 \label{eq:GeneralExpect1}
\end{eqnarray}
Thus, we can interchange scalar values to reform the expectation value of $\hat{Z}$ for qubit $C$ in the form 
\begin{eqnarray} 
{}_C \bra{\Psi^4 } \hat{Z} \ket{\Psi^4 } {}_{C} && = P_0 - P_1  \equiv \langle \Phi | \hat{Z}^{\chi}_{A} | \Phi \rangle \;,
\end{eqnarray}
where the new operator $\hat{Z}^{\chi}_{A}$ with $\hat{\rho} (t) = \ket{\psi (t)}\bra{\psi (t)}$ is given by
\begin{eqnarray} 
\hat{Z}^{\chi}_{A} = {1\over 2} \Big( e^{i \chi} \hat{N}\, \hat{\rho} (t) \hat{M}  \hat{A}  + e^{-i \chi}   \hat{A}^{\dag} \hat{M}^{\dag}\, \hat{\rho} (t) \hat{N}^{\dag}\, \Big)\, .
 \label{eq:GeneralZ01-2}
\end{eqnarray}

For example, choosing $\chi = 0$ and $\hat{A} = \openone$, this expression generates 
\begin{eqnarray} 
\langle \Phi| \hat{Z}_1^{0}|\Phi \rangle =  {1\over 2} \Big(  \bra{\Phi}\hat{N}\, \hat{\rho} (t) \hat{M}\, \ket{\Phi} +  \bra{\Phi} \hat{M}^{\dag}\, \hat{\rho} (t) \hat{N}^{\dag}\, \ket{\Phi} \Big) =   \left< {\Phi} \left| {1\over 2} \left( \hat{N}\, \hat{\rho} (t) \hat{M} + \hat{M}^{\dag}\, \hat{\rho} (t) \hat{N}^{\dag}\right) \right| {\Phi} \right>,
\label{eq:GeneralZ02}
\end{eqnarray}
whilst with $\chi = \pi/2$ and $\hat{A} = \openone$, instead it generates
\begin{eqnarray} 
\langle \Phi| \hat{Z}_1^{\pi/2}| \Phi \rangle =  {i\over 2} \Big(  \bra{\Phi}\hat{N}\, \hat{\rho} (t) \hat{M}\, \ket{\Phi} -   \bra{\Phi} \hat{M}^{\dag}\, \hat{\rho} (t) \hat{N}^{\dag}\, \ket{\Phi} \Big) =  \left< {\Phi} \left| {i \over 2} \left( \hat{N}\, \hat{\rho} (t) \hat{M} - \hat{M}^{\dag}\, \hat{\rho} (t) \hat{N}^{\dag}\right) \right| {\Phi} \right>.
\label{eq:GeneralZ03}
\end{eqnarray}

\subsection{A brief derivation of the Lindblad master equation at $t = 2 \delta t$}
\label{Lindblad-derivation}
In the concept of discretised time evolution with a step of $\delta t$, the initial density matrix is taken as $\hat{\rho} (0) = \ket{\psi_0}\bra{\psi_0}$.
After time $\delta t$, $\hat{\rho} (0)$ follows a unitary time evolution and becomes $\hat{\rho} (\delta t) = \hat{U} (\delta t) \hat{\rho} (0) \left( \hat{U} (\delta t)\right)^{\dag} $, where the unitary operator is given by
\begin{eqnarray} 
\hat{U} (\delta t) = \sum_{j=0}^{\infty} {(-i\, \delta t)^j \over j!} \left( \hat{\cal H} \right)^j = \openone - i \,  \hat{\cal H} \,\delta t - {1 \over 2} \hat{\cal H}^2 \, (\delta t)^2 + ...\, .
\label{Unitary-deriv01} 
\end{eqnarray}
Now, we take into account the decoherence in the time evolution step from $\delta t$ to $2 \delta t$, to describe an open quantum system. We adopt the Kraus representation of quantum operations for single Lindblad operator $\hat{\cal L}$ given by
\begin{eqnarray} 
&& \hat{\rho} (2\delta t) = \sum_{r=0,1} \hat{\cal N}_r \, \hat{\rho} (\delta t) ~ \hat{\cal N}^{\dag}_r \; ,
\label{Lind-deriv01} \\
&& \sum_{r=0,1}  \hat{\cal N}^{\dag}_r \, \hat{\cal N}_r \approx \openone + O (\delta t^2)\, ,
\label{Lind-deriv02} 
\end{eqnarray}
for $\hat{\cal N}_0 = \openone + \left(-i \hat{\cal H}- {1 \over 2} \hat{\cal L}^{\dag}  \hat{\cal L} \right) \delta t$, $\hat{\cal N}_1 = \sqrt{\delta t} \hat{\cal L} $ and $O (\delta t^2) = \left( \hat{\cal H}^2 - {i \over 2} \hat{\cal H} \hat{\cal L}^{\dag}  \hat{\cal L} 
+ {i \over 2} \hat{\cal L}^{\dag}  \hat{\cal L} \hat{\cal H}- {1 \over 2} \hat{\cal L}^{\dag}  \hat{\cal L} \hat{\cal L}^{\dag}  \hat{\cal L} \right) (\delta t )^2$.
Thus, the density matrix at $t=2\delta t$ can be rewritten as
\begin{eqnarray} 
&& \hat{\rho} (2\delta t) \approx \hat{\rho} (\delta t) + \left( i \left[ \hat{\rho}(\delta t) , \hat{\cal{H}} \right] + \hat{\cal L } \hat{\rho}(\delta t)\, \hat{\cal L }^{\dag} - {1\over 2} \left\{ \hat{\rho} (\delta t)\, , \hat{\cal L }^{\dag} \hat{\cal L } \right\} \right) \delta t + Q (\delta t^2)\, ,
\label{Lind-deriv03} 
\end{eqnarray}
where
$Q (\delta t^2) = \left( \hat{\cal H} \hat{\rho} (\delta t)  \hat{\cal H} + {i \over 2} \hat{\cal H} \, \hat{\rho} (\delta t)\, \hat{\cal L}^{\dag}  \hat{\cal L} 
- {i \over 2} \hat{\cal L}^{\dag}  \hat{\cal L} \, \hat{\rho} (\delta t)\, \hat{\cal H}- {1 \over 2} \hat{\cal L}^{\dag}  \hat{\cal L} \, \hat{\rho} (\delta t)\, \hat{\cal L}^{\dag}  \hat{\cal L} \right) (\delta t )^2 $.
Finally, taking the limit $\delta t \rightarrow 0$ generates the Lindblad equation (\ref{LindEq01}).

\section*{Author contributions}
J. J. conceived the idea and both authors discussed the results together and contributed to the writing and theory development for the final manuscript.

\section*{Competing interests}
The authors declare no competing interests.


\section*{References}
\begin{thebibliography}{}

\bibitem{Quantum-phys-01}
Born, M. \& Jordan, P. The 1925 Born and Jordan paper “On quantum mechanics” {\em Z. Phys} {\bf 34}, 858, (1925)

\bibitem{Quantum-phys-02}
Dirac, P. A. M. The fundamental equations of quantum mechanics {\em Proc. R. Soc. London, Ser. A} {\bf 109}, 642 (1925).

\bibitem{Schroedinger01}
Schrödinger, E. An undulatory theory of the mechanics of atoms and molecules {\em Phys. Rev.} {\bf 28}, 1049 (1926).

\bibitem{VNE01}
von Neumann, J. {\em Göttinger Nachrichten} {\bf 245} (1927).

\bibitem{Peter-RMP}
Plenio M. B. \& Knight, P. L. {\em Rev. Mod. Phys.} {\bf 70}, 101 (1998).

\bibitem{Lindblad01}
Lindblad, G. On the generators of quantum dynamical semigroups {\em Commun. Math. Phys.} {\bf 48}, 119 (1976).

\bibitem{Lindblad02}
Gorini, V. Kossakowski, A. \& Sudarshan, E. C. G. Completely positive dynamical semigroups of N‐level systems {\em Jour. of Math. Phys.} {\bf 17}, 821 (1976).

\bibitem{SethLloyd96} 
Lloyd, S. Universal quantum simulators {\em Science} {\bf 273}, 1073 (1996).

\bibitem{PRXQuantum22}
Kamakari, H. Sun, S.-N. Motta, M. \& Minnich, A. J.  
Digital quantum simulation of open quantum systems using quantum imaginary–time evolution {\em PRX Quantum} {\bf 3}, 010320 (2022).

\bibitem{Cleve17}
Cleve R. \& Wang, C. Efficient quantum algorithms for simulating lindblad evolution
{\em 44th International Colloquium on Automata, Languages, and Programming} {\bf 17} (2017).

\bibitem{SBen20}
Endo, S. Sun, J. Li, Y. Benjamin, S. C. \& Yuan, X.  
Variational quantum simulation of general processes 
{\em Phys. Rev. Lett.} {\bf 125}, 010501 (2020).

\bibitem{deJong21}
Metcalf, M. Stone, E. Klymko, K. Kemper, A. F. Sarovar, M. \& de Jong, W. A.  
Quantum Markov chain Monte Carlo with digital dissipative dynamics on quantum computers
arXiv:2103.03207.

\bibitem{Childs17}
Childs, A. M. \& Li, T.  
Efficient simulation of sparse Markovian quantum dynamics
{\em Quan. Inf. and Comp.} {\bf 17}, 901 (2017).

\bibitem{Eisert11}
Kliesch, M. Barthel, T. Gogolin, C. Kastoryano, M. \& Eisert, J. 
Dissipative quantum Church-Turing theorem
{\em Phys. Rev. Lett.} {\bf 107} 120501, 2011.

\bibitem{Cirac09}
 Verstraete, F. Wolf, M. M. \& Cirac, J. I.  
Quantum computation and quantum-state engineering driven by dissipation
{\em Nature Physics} {\bf 5} 633 (2009).

\bibitem{JJ2021}
Joo, J. \& Moon, H. Quantum variational PDE solver with machine learning arXiv:2109.09216.

\bibitem{JJ2020}
Lubasch, M. Joo, J. Moinier, P. Kiffner, M. \& Jaksch, D.  
Variational quantum algorithms for nonlinear problems
{\em Phys. Rev. A} {\bf 101}, 010301(R) (2020).

\bibitem{Note1}
Note that one can rewrite the same Lindblad equation only using commutation relations alternatively.

\bibitem{Saito19}
Funo, K. Shiraishi, N. \& Saito, K.  
Speed limit for open quantum systems
{\em New J. Phys.} {\bf 21}, 013006 (2019).

\bibitem{Hasegawa21}
Van Vu, T. \& Hasegawa, Y.  
Lower bound on irreversibility in thermal relaxation of open quantum systems
{\em Phys. Rev. Lett.} {\bf 127}, 190601 (2021).

\end{thebibliography}
\end{document}